\documentclass[lettersize,journal]{IEEEtran}
\usepackage{amsmath,amsfonts}
\usepackage{algorithmic}
\usepackage{algorithm}
\usepackage{array}
\usepackage[caption=false,font=normalsize,labelfont=sf,textfont=sf]{subfig}
\usepackage{textcomp}
\usepackage{stfloats}
\usepackage{url}
\usepackage{verbatim}
\usepackage{graphicx}
\usepackage{cite}
\usepackage{amssymb}
\usepackage{xcolor}
\usepackage{framed}
\usepackage{multirow}
\usepackage{multicol}
\usepackage{lipsum, booktabs}  
\usepackage{algorithm}
\usepackage{algorithmic}

\newtheorem{prot}{Protocol}
\newcommand{\BPR}{\begin{prot}}   \newcommand{\EPR}{\end{prot}}
\newtheorem{theorem}{Theorem}[section]
\newtheorem{lemma}{Lemma}[section]

\newtheorem{definition}[theorem]{Definition}

\newtheorem{proposition}[theorem]{Proposition}

\newtheorem{rule-def}[theorem]{Rule}
\numberwithin{equation}{section} \makeatletter
\newsavebox{\savepar}

\begin{document}

\title{An Undeniable Signature Scheme Utilizing Module Lattices}
\author{Kunal Dey, Mansi Goyal, Bupendra Singh and Aditi Kar Gangopadhyay 
\thanks{Kunal Day is with the Department of Computer Science, University of Calgary, Alberta, Canada (e-mail: kunaldey3@gmail.com)}
\thanks{Mansi Goyal is with the Department of Mathematics, Indian Institute of Technology Roorkee, Uttarakhand, India (e-mail: mansi@ma.iitr.ac.in),}
\thanks{Bupendra Singh is Scientist F, CAIR, DRDO, Bengaluru, India (e-mail: bhusinghdrdo@gmail.com),}
\thanks{ Aditi Kar Gangopadhyay is with the Department of Mathematics, Indian Institute of Technology Roorkee, Uttarakhand, India (e-mail: aditi.gangopadhyay@ma.iitr.ac.in)}

}

\maketitle

\begin{abstract}

An undeniable signature scheme is type of digital signature where the signer retains control over the signature's verifiability. Therefore with the approval of the signer, only an authenticated verifier can verify the signature. In this work, we develop a module lattice-based post-quantum undeniable signature system. Our method is based on the GPV framework utilizing module lattices, with the security assured by the hardness of the {\sf SIS} and {\sf LWE} problems.
%, as outlined in \cite{bert2021implementation} where the security follows from the hardness of the {\sf SIS} and {\sf LWE} problems.
We have thoroughly proved all the desired securities for the proposed scheme. Finally, we have implemented our protocol for different sets of parameters. The purpose of opting a module variant rather than a ring variant is to provide greater flexibility in selecting parameters.
\end{abstract}

\begin{IEEEkeywords} Digital signature, Lattice-based cryptography, Module lattices, Post-quantum cryptography, {\sf LWE} problem, {\sf SIS} problem, Undeniable signature.
\end{IEEEkeywords}

\section{Introduction}
\IEEEPARstart{A} digital signature is a cryptographic mechanism used to authenticate messages. It typically involves three algorithms: one for generating the user's public and secret keys, a signing algorithm, and a verification algorithm. Unlike traditional digital signatures, undeniable signatures require the active involvement of the signer to verify the signature. This concept was first introduced by Chaum and van Antwerpen \cite{chaum1990undeniable}. Unlike standard digital signatures, which can be verified by anyone with access to the signer's public key, undeniable signatures require an interactive protocol with the signer for verification. This design grants the signer control over who is permitted to verify their signatures, providing an added layer of selective verification. During this protocol, the verifier obtains either an affirmative (confirmation protocol) or negative response (disavowal protocol), indicating whether the signature generated by the original signer. The signer is unable to deceive the verifier by making a valid signature appear invalid or an invalid signature appear valid. Given that merely copying a signature does not yield directly verifiable results, undeniable signatures are favored over conventional digital signatures for specific applications such as e-voting, e-cash, etc. Over years, there has been notable research activity focused on undeniable signature schemes \cite{boyar1991convertible, damgaard1996new, gennaro2000rsa, galbraith2002rsa, schuldt2010efficient, zhao2012certificateless}. These cryptographic schemes are based on number-theoretic problems, such as the discrete logarithm and integer factorization problems, which are vulnerable to quantum attacks due to {\em Shor's algorithm} \cite{shor1999polynomial}. This poses a significant threat to modern cryptographic protocols, as many widely adopted encryption, key exchange, and digital signature schemes could become obsolete with the advent of powerful quantum computers. This urgency underscores the need to transition toward quantum-safe cryptographic methods. Post-quantum cryptography (PQC) evolved to provide quantum-resistant protocols in response to the growing threat of quantum computing. Several candidates for PQC exist, including  {\em Lattice-based cryptography} \cite{ravi2021lattice}, {\em Isogeny-based cryptography} \cite{de2017mathematics}, {\em Code-based cryptography} \cite{weger2022survey}, {\em Hash-based cryptography} \cite{srivastava2023overview}, {\em Multivariate cryptography} \cite{ding2009multivariate}. Each of these PQC approaches has distinct advantages and disadvantages, making direct comparisons challenging. Numerous post-quantum undeniable signature schemes have been proposed in the literature on PQC. Jao et al. \cite{jao2014isogeny} introduced an isogeny-based undeniable signature scheme, and Srikant et al. \cite{srinath2016isogeny} extended it to a blind signature version. Additionally, there are code-based undeniable signature schemes \cite{aguilar2013code,hua2018undeniable}. On the other hand, a lattice-based undeniable signature scheme has been proposed, by Rawal et al. \cite{rawal2022lattice}.  

Lattice-based cryptography being one of the most promising candidates in the area of PQC. In 2016, the post-quantum cryptography standardisation initiative was started by the National Institute of Standards and Technology (NIST) aiming to update its cryptographic standards with post-quantum cryptography to safeguard against the threat posed by quantum computers. This project launched a global competition to identify quantum-resistant cryptographic algorithms capable of replacing classical schemes vulnerable to quantum attacks. By 2023, NIST had selected three algorithms for post-quantum digital signatures. The noteworthy aspect is that two of these algorithms are lattice-based: CRYSTALS-Dilithium \cite{ducas2018crystals, lyubashevsky2020crystals} and FALCON \cite{fouque2018falcon}. While CRYSTALS-Dilithium employs the Schnorr signature framework \cite{seurin2012exact}, FALCON follows the GPV signature framework \cite{gentry2008trapdoors}. Over the years, various variants derived from the Schnorr and GPV frameworks have been proposed, resulting in advancements in cryptographic schemes. These include group signatures \cite{zhang2021survey, gordon2010group}, which facilitate anonymous authentication within a group; ring signatures \cite{cayrel2010lattice, lu2019raptor}, which maintain signer anonymity without disclosing the identity of the member who signed a message; and one undeniable signature \cite{rawal2022lattice}, which require the signer's cooperation for verification. These innovations highlight the adaptability and potential of lattice-based cryptography in addressing the challenges posed by quantum computing.

The initial preference for lattice-based constructions often centers on ideal lattices and ring settings \cite{lyubashevsky2010ideal, stehle2009efficient, peikert2006efficient}. However, module lattices \cite{langlois2015worst} offer a compelling alternative, as schemes based on a module setting are nearly as efficient as their ring-based counterparts while providing additional practical advantages, including computational efficiency and low memory usage \cite{bert2021implementation}. Furthermore, module lattice-based schemes can potentially reduce the sizes of public keys and ciphertexts while ensuring robust security guarantees, making them particularly suitable for constrained environments such as embedded systems or mobile devices. Notably, the NIST-standardized CRYSTALS-Dilithium scheme \cite{ducas2018crystals} is based on module lattices, and a recent study \cite{chuengsatiansup2020modfalcon} introduces a module version of the Falcon signature scheme. In light of these advantages, module lattices are increasingly being viewed as a promising alternative for post-quantum cryptography, potentially offering a more scalable and secure foundation for future cryptographic schemes. Despite the advantages of module lattices, there is a notable absence of undeniable signature schemes based on this framework. Therefore, developing such a protocol presents an intriguing research opportunity.

\subsection{Our Contribution}
Ideal lattices, often constructed from rings of the type $\mathcal{R}_q = \mathbb{Z}_q [X] / (X^n + 1)$, are commonly preferred for designing efficient lattice-based schemes. Module lattices \cite{langlois2015worst}, which use modules represented by $\mathcal{R}_q^d$, provide a balance between ideal and unstructured lattices. The key contributions made by this work are summarized below.
\begin{itemize}
    \item In this work, we build upon module lattices and propose an undeniable signature scheme based on the GPV signature \cite{bert2021implementation,gentry2008trapdoors} over module lattices. Our scheme's security relies on the {\sf module-LWE} and {\sf module-SIS} problems. The primary goal of our work is to design a module-based variant of the undeniable signature scheme that provides enhanced flexibility in parameter selection rather than focusing solely on performance.
    \item Rawal et al. \cite{rawal2022lattice} introduced an undeniable signature scheme using the ring setting. In their construction, a dishonest signer with a valid signature could successfully execute the {\sf Disavowal} protocol with a non-negligible advantage, which contradicts the fundamental properties of an undeniable signature scheme. In contrast, our construction prevents a dishonest signer from running the {\sf disavowal} protocol with a valid signature with a non-negligible probability. The {\sf disavowal} 
 and {\sf confirmation} protocols of our scheme is influenced by the Stern's identification scheme \cite{stern1996new, kawachi2008concurrently}.
    \item We leverage the work on trapdoors over module lattices by Bert et al. \cite{bert2021implementation}, who implemented lattice trapdoors for various security parameters. Our protocol was implemented in C, and we evaluated the communication and computation costs using proposed parameters for different security levels.
\end{itemize}
%-------------------------------
\subsection{Organization of the paper}
The structure of this paper is organized as follows: Section \ref{Preliminaries} outlines the essential preliminaries needed for a comprehensive understanding of our scheme. Section \ref{Proposed}, introduces our undeniable signature scheme, which is built on module lattices. Section \ref{Security} discusses the security analysis, while Section \ref{complexity} evaluates the communication and computation costs. Lastly, Section \ref{Conclusion} provides a summary of the key findings of the paper.
%------------------------
\section{Preliminaries}\label{Preliminaries}
\noindent \textbf{Notations}. Bold lowercase letters (e.g., {\bf a}, {\bf h}) are used to denote vectors, while bold uppercase letters (e.g., {\bf U}, {\bf E}) represent matrices. The Euclidean norm is denoted by $\parallel \cdot \parallel$, and for a matrix $\mathbf{X}$, the norm is defined as $\parallel \mathbf{X} \parallel = \max_i \parallel \mathbf{x}_i \parallel$, where $\mathbf{x}_i$ is the ith column of $\mathbf{X}$. The norm of a vector in $\mathbb{Z}_q$ is defined as the norm of the corresponding vector over $\mathbb{Z}$ with each entry taken from the set $ \{-\lfloor \frac{q}{2} \rfloor, \ldots, \lfloor \frac{q}{2} \rfloor\} $. The symbol $\cdot \parallel \cdot$ represents the concatenation of two entities. $\mathcal{U}(S)$ represents the uniform distribution over the finite set $S$. $\omega(\cdot)$ denotes the lower bound of the asymptotic growth rate. Let ${\bf u}$ and ${\bf v}$ be two vectors in $\mathbb{R}^{n}$. The expression  $<{\bf u}, {\bf v}>$ denotes the standard dot product, defined by $<{\bf u}, {\bf v}> = {\bf u}\cdot{\bf v}^{T}$.

\begin{definition}{\textbf{Negligible function}}:\\
A function ${\sf negl}:\mathbb{N} \longrightarrow \mathbb{R}$ is considered negligible if it decreases faster than the inverse of any polynomial.

% for every polynomial $\text{poly}(x)$, there is a positive integer $N$ such that ${\sf negl}(x) < \frac{1}{\text{poly}(x)}$ for all \( x > N \).
\end{definition}

\begin{definition}{\textbf{Lattice}.}
Given $m$ linearly independent vectors $\mathbf{p}_1, \mathbf{p}_2, \ldots, \mathbf{p}_m \in \mathbb{R}^n$, the {\em lattice} generated by these vectors is defined as follows:
\begin{equation*}
    \mathcal{L} = \mathcal{L}(\mathbf{p}_1, \mathbf{p}_2, \ldots, \mathbf{p}_m) = \left\{ \sum_{i=1}^{m} x_i \mathbf{p}_i \mid x_i \in \mathbb{Z} \right\}.
\end{equation*}
The set $\{\mathbf{p}_1, \ldots, \mathbf{p}_m\} $ forms a basis for $\mathcal{L}$. Here, $m$ denotes the rank, and $n$ represents the lattice's dimension. When $n \text{equals} m$, the lattice is referred to as a full-rank lattice.
\end{definition}

\noindent The {\em dual lattice} $\mathcal{L}^{*}$ corresponding to $\mathcal{L}$ is defined by,
\begin{equation*}
    \mathcal{L}^{*} = \{\mathbf{w}\in span(\mathcal{L}): \forall \mathbf{v}\in \mathcal{L}, <\mathbf{w}, \mathbf{v}> \in \mathbb{Z}\}.
\end{equation*}
%%%%%%%%%%%%%%%%%%%%%%%%%%%%%%%%%%%%%%%%%%%%%%%%%%%%%%%%%%%%%%%%%%%%%%%%%%%%%%%%%%%%%%%%%%%%%%%%%%%%%%%%%%%%%%%%%%%%%%%%%%%%%%%
\subsection{Gaussian on Lattices \cite{gentry2008trapdoors, chi2015lattice}}
Consider a basis $B \subset \mathbb{R}^n$ for an n-dimensional lattice $\mathcal{L}$ (or a full-rank set $S \subset \mathcal{L} $). The Gaussian sampling algorithm randomly selects a lattice vector according to a probability distribution resembling the Gaussian distribution. 

\begin{definition}{\textbf{Gaussian Function on $\mathbb{R}^n$}.} Let $\mathcal{L}$ be an n-dimensional lattice, $\mathbf{c} \in \mathbb{R}^n$, and $s > 0$. A Gaussian function centered at $\mathbf{c}$ on $\mathbb{R}^n$ is defined as follows:
\begin{align*}
    \rho_{s,\mathbf{c}}(\mathbf{x}) = \exp \left( \frac{-\pi \|\mathbf{x} - \mathbf{c}\|^2}{s^2} \right), ~ \forall ~ \mathbf{x} \in \mathbb{R}^n.
\end{align*}
If $s = 1$ and $\mathbf{c} = \mathbf{0}$, we can eliminate the subscripts $s$ and $\mathbf{c}$.
\end{definition}

\begin{definition}{\textbf{Discrete Gaussian Distribution}.} Let $\mathcal{L}$ be an n-dimensional lattice, $s > 0$ and $\mathbf{c} \in \mathbb{R}^n$. The discrete Gaussian distribution over $\mathcal{L}$ is defined as:
\begin{align*}
    D_{\mathcal{L}, s, \mathbf{c}}(\mathbf{x}) = \frac{\rho_{s,\mathbf{c}}(\mathbf{x})}{\rho_{s,\mathbf{c}}(\mathcal{L})}, ~ \forall ~ \mathbf{x} \in \mathcal{L},
\end{align*}
where $\rho_{s,\mathbf{c}}(\mathcal{L}) = \sum_{\mathbf{y} \in \mathcal{L}} \rho_{s,\mathbf{c}}(\mathbf{y})$.
\end{definition}
%---------------------------------------------------------------
\subsection{Some Hard Problems in Lattice}
Ajtai \cite{ajtai1996generating} was the first to introduce the hard-in-average problem, which involves finding a short nonzero vector $\mathbf{s} \in \mathbb{Z}^n$ for a linear system $\mathbf{A} \mathbf{s} = 0 \mod q$, where $\mathbf{A} \in \mathbb{Z}_q^{m \times n}$ is randomly generated. This problem is commonly referred to as the Short Integer Solution ({\sf SIS}) problem, while its inhomogeneous variant is called the Inhomogeneous Short Integer Solution ({\sf ISIS}) problem.

\begin{enumerate}
   
    \item {\textbf{The Shortest Integer Solution Problem}({\sf SIS}):} Let we have a matrix $\mathbf{A} \in \mathbb{Z}_q^{m \times n}$, a real number $\beta$ and an integer $q$. The problem is to find a non-zero vector $\mathbf{s} \in \mathbb{Z}^n$ such that $\mathbf{A} \mathbf{s} = 0 \mod q$ with $\|\mathbf{s}\| \le \beta$.

    \item {\textbf{Inhomogeneous Shortest Integer Solution Problem}({\sf ISIS}):} Let we have a matrix $\mathbf{A} \in \mathbb{Z}_q^{m \times n}$, an integer $q$ and a real number $\beta$ and $\mathbf{u} \in \mathbb{Z}_q^m$. The problem is to find a non-zero vector $\mathbf{s} \in \mathbb{Z}^n$ such that $\mathbf{A}\mathbf{s} = \mathbf{u} \mod q$ with $\|\mathbf{s}\| \le \beta$.

    \item {\textbf{Learning with Error }({\sf LWE}) \cite{regev2009lattices}} \\  {\sf LWE} is a computational problem where secret information is embedded into a system of linear equations, with small errors added to perturb the system.\\%involves encoding secret information into a system of linear equations that are perturbed by errors.
  {\textbf{Learning with Error}({\sf LWE})(Decisional version):} \\ 
    Suppose $\mathcal{E}$ is an error distribution over $\mathbb{Z}_q$ for an integer $q$ and $n$ is a positive integer. Let the probability distribution $\mathcal{X}_{\mathbf{s},\mathcal{E}}$ on $\mathbb{Z}_q^n \times \mathbb{Z}_q$ be defined as:\\
    Choose a vector ${\mathbf{a}}$ from $\mathbb{Z}_q^n$ uniformly, and $x$ according to $\mathcal{E}$, and for ${\mathbf{s}} \in \mathbb{Z}_q^n$, output $(\mathbf{a}, \mathbf{a}^{T} \mathbf{s}+{x})$.\\ The goal of this problem is to distinguish between the distribution $\mathcal{X}_{\mathbf{s},\mathcal{E}}$ for some $ {\mathbf{s}} \in \mathbb{Z}_q^n$ and the uniform distribution of $\mathbb{Z}_q^n \times \mathbb{Z}_q$. Note that all of the above operations are done in $\mathbb{Z}_q$. 
\end{enumerate}
%------------------------------------------------------------------
\subsection{Hard Random Lattices}
Let $n,m,q$ be some positive integers, where $n$ be the security parameter and all other variables are functions of $n$ and $\mathbf{A}\in \mathbb{Z}_q^{n\times m}$. We now define an $m$-dimensional lattice of full rank, denoted by $\mathcal{L}^{\perp}(\mathbf{A})$.
\begin{align*}
      \mathcal{L}^{\perp}(\mathbf{A})=\{\mathbf{e}\in \mathbb{Z}^m : \mathbf{A}\mathbf{e}=\mathbf{0} \mod q\}.
\end{align*}
The set of syndromes are defined as,
\begin{align*}
synd = \{\mathbf{u}=\mathbf{A}\mathbf{e}\mod{q} : \mathbf{e} \in \mathbb{Z}^m\}.
\end{align*}
Here $synd$ is a subset of $\mathbb{Z}_q^n$. It can be shown that $synd \approx {\mathbb{Z}^m/ \mathcal{L}^{\perp}(\mathbf{A}) }$.\\

The following Lemma \ref{Lemma 1} from \cite{gentry2008trapdoors} is instrumental in defining a trapdoor function, as discussed in the subsequent subsection. Readers are encouraged to consult \cite{gentry2008trapdoors, micciancio2012trapdoors, genise2018faster} for the essential background on sampling from the discrete Gaussian distribution.

\begin{lemma}\label{Lemma 1} Let $m \geq 2n \log q$, where $q$ is a prime number and $n$ is a positive integer. Then, for all $s \geq \omega(\sqrt{\log m}) $ and for any but a $ 2 q^{-n} $ fraction of matrices $\mathbf{A} \in \mathbb{Z}_q^{n \times m} $, the following holds:
\begin{enumerate}
\item There exists a vector $\mathbf{e} \in \{0,1\}^m$ such that $\mathbf{A}\mathbf{e} = \mathbf{u} \mod{q}$ for every syndrome $\mathbf{u} \in \mathbb{Z}_q^n$.

\item The distribution of $\mathbf{u} = \mathbf{A}\mathbf{e} \mod{q}$ is statistically close to uniform across $\mathbb{Z}_q^n$ for $\mathbf{e} \sim D_{\mathbb{Z}^m,s}$.

\item The conditional distribution of $\mathbf{e} \sim D_{\mathbb{Z}^m,s}$, given $\mathbf{u} = \mathbf{A}\mathbf{e} \mod{q}$, is precisely $\mathbf{t} + D_{\mathcal{L}^{\perp}(\mathbf{A}),s,-\mathbf{t}}$. This is for a fixed $\mathbf{u} \in \mathbb{Z}_q^n$ and an arbitrary solution $\mathbf{t} \in \mathbb{Z}^m$ to the equation $\mathbf{u} = \mathbf{A}\mathbf{t} \mod{q}$.
\end{enumerate}
\end{lemma}
%---------------------------------------------
\subsection{Trapdoor Functions}
\label{Trapd}
A {\em trapdoor function} \cite{yao1982theory, gentry2008trapdoors, micciancio2012trapdoors} refers to a type of mathematical function that is simple to evaluate in one direction but becomes hard to reverse without access to a specific key, known as the {\em trapdoor}. The trapdoor is typically a secret key that is used to enable efficient computation of the function in the reverse direction. Trapdoor functions play a crucial role in lattice-based cryptography, where they are used to produce secret key and public key pair for both encryption and digital signatures schemes.
%In the context of lattice-based cryptography, trapdoor functions are used to generate public and secret keys for signature and encryption schemes. 
In a typical lattice-based signature scheme, the signer generates a random lattice and applies a trapdoor function to it to obtain a public key. The trapdoor enables the signer to efficiently generate a secret key that can be used to sign messages. Trapdoor functions with preimage sampling are determined by three {\sf PPT} algorithms $(\text{TrapGen, SampleDom, SamplePre})$.
\begin{itemize}
\item $(a,t) \longleftarrow \text{TrapGen}(1^n)$: Given the security parameter this algorithm outputs the pair $(a,t)$, where $a$ is the description of a function $f_a:D_n \longrightarrow R_n$ and $t$ is a trapdoor for the function $f_a$.
\item $x \longleftarrow\text{SampleDom}(1^n)$: Given the security parameter, this algorithm samples $x$ from $D_n$ in such a way that $f_a(x)$ is uniformly distributed over $R_n$.
\item $x \longleftarrow\text{SamplePre}(t,y)$: For every $y\in R_n$ this algorithm sample from a conditional distribution of $x \leftarrow$SampleDom$(1^n)$ such that $f_a(x)=y$.
\end{itemize}

\noindent To proceed with construction, we refer to the result by Ajtai \cite{ajtai1999generating}, which demonstrates a method for sampling an approximately uniform matrix $\mathbf{A}\in \mathbb{Z}_q^{n\times m}$ and simultaneously generating a short, full-rank trapdoor set of lattice vectors $\mathbf{S}\subseteq \mathcal{L}^{\perp}(\mathbf{A})$.

\begin{proposition}\label{prop 1} Let q be a prime number, and let m and n be two positive integers such that $m \geq 5n \log q$. Then, there exists a probabilistic polynomial time {\sf (PPT)} algorithm that outputs a matrix $\mathbf{A} \in \mathbb{Z}_q^{n \times m}$ and a full-rank set $\mathbf{T} \subseteq \mathcal{L}^{\perp}(\mathbf{A})$ based on the input security parameter n. The distribution of $\mathbf{A}$ is uniform over $\mathbb{Z}_q^{n \times m}$, and the norm $\|\mathbf{T}\| \leq L = m^{2.5}$.
\end{proposition}
\vspace{0.2cm}
%------------------------------------------
\subsubsection*{Construction of Trapdoor Function Using {\sf SIS}}
Let $q, m$, and $L$ be as defined in Proposition \ref{prop 1}. The collection of trapdoor functions with preimage sampling are determined by a Gaussian parameter $s \geq L \cdot \omega(\sqrt{\log m}) $. The construction is described as follows.
\begin{enumerate}
\item The pair $(\mathbf{A},\mathbf{T})$ is selected based on the criteria outlined in the Proposition \ref{prop 1}.
\item The matrix $\mathbf{A}$ defines a function $f_{\mathbf{A}}$ such that $f_{\mathbf{A}} (\mathbf{e})=\mathbf{A} \mathbf{e} \mod{q}$, where $D_n=\{\mathbf{e}\in \mathbb{Z}^m : \|\mathbf{e}\| \leq s \sqrt{m}\}$ is the domain and $R_n=\mathbb{Z}_q^n$ is the the range. 
\item Given a tuple $(\mathbf{A}, \mathbf{T}, s, \mathbf{u})$, the task is to find a solution to the equation $\mathbf{A}\mathbf{t} = \mathbf{u} \mod{q}$. According to Lemma \ref{Lemma 1}, such a vector $\mathbf{t}$ can be selected. Once this is done, sample $\mathbf{v}$ from the distribution $D_{\mathcal{L}^{\perp}(\mathbf{A}),s,-\mathbf{t}}$ and then produce the solution $\mathbf{e} = \mathbf{t} + \mathbf{v}$.
\end{enumerate}

\begin{theorem}\label{Theorem 2}
If the ${\sf ISIS}_{q,m,\beta = 2s\sqrt{m}}$ problem is hard, then the construction described above will form a collection of trapdoor functions.
\end{theorem}

\subsection{The GPV Signature}

The GPV signature scheme, introduced by Gentry, Peikert, and Vaikuntanathan \cite{gentry2008trapdoors}, is a lattice-based cryptographic signature scheme. It serves as a modified version of the traditional {\sf Fiat-Shamir} signature scheme and depends on the computational difficulty of the {\em Short Integer Solution} ({\sf SIS}) problem in lattice-based cryptography.

\noindent\textbf{Construction}:
Let $n\in \mathbb{N}$ be a security parameter. A {\em collision-resistant} hash function $H=H_n:\{0,1\}^{*} \longrightarrow R_n$ has been used to model it as a random oracle. The GPV signature has three algorithms $\text{KeyGen}$, $\text{Sign}$ and $\text{Verification}$ as discussed below:

\begin{itemize}
\item[] $(\text{pk,sk}) \leftarrow \text{KeyGen}(1^n)$: 
\begin{enumerate}
    \item The signer runs the $\text{TrapGen}$ algorithm and receives the pair $(a,t)$, where $t$ is its trapdoor and $a$ is the description of the function $f_a$ (see Subsection \ref{Trapd}).
    \item Public key $\text{pk}=a$ and secret key $\text{sk}=t$.
\end{enumerate}

\item[] $(\sigma) \leftarrow \text{Sign}(\text{sk},msg)$: 
\begin{enumerate}
    \item Parse $\text{sk}= t$.
    \item The signer computes $H(msg)$.
    \item Generates $\sigma \leftarrow \text{SamplePre}(t, H(msg))$. 
\end{enumerate}

\item[] $(\top \text{or} \bot) \leftarrow \text{Verification}(msg, \text{a},\sigma)$: 
\begin{enumerate}
    \item Parse $\text{pk}= a$.
    \item If $\sigma \not\in D_n$ output $\bot$.
    \item Else computes $f_a(\sigma)$.
    \item If $f_a(\sigma) = H(msg)$, output $\top$ else output $\bot$.
\end{enumerate}
\end{itemize}

\begin{theorem}
Based on the belief that the Short Integer Solution $({\sf SIS})$ problem is difficult to solve, the GPV scheme guarantees existential unforgeability against chosen message attacks (EUF-CMA).
\end{theorem}

%----------------------------------------------------------------------

\subsection{Undeniable Signature Scheme}

In our work, we adhere to the definition of undeniable signatures as outlined in \cite{kurosawa2008universally, jao2014isogeny}.

\begin{definition}{(Undeniable signature).} An undeniable signature scheme is a tuple of polynomial-time algorithms $({\sf KeyGen}, {\sf Sign}, {\sf Confirmation}, {\sf Disavowal})$. Let $n\in \mathbb{N}$ be a security parameter. This scheme entails interactions between the signer (S) and the verifier (V).
\begin{enumerate}
\item $(\text{pk,sk}) \leftarrow{\sf KeyGen}(1^{n})$: Given the security parameter $n$ as input, the algorithm produces a signing key $\text{sk}$ and a verification key (public key) $\text{pk}$.
\item $\sigma \leftarrow {\sf Sign}(\text{msg, sk})$: This algorithm takes $(\text{msg, sk})$ as inputs and generates a signature $\sigma$.
\item $(0/1) \leftarrow {\sf Verification}(\sigma,\text{msg})$: It consists two subprotocols:

\begin{itemize}
    \item $(0/1) \leftarrow {\sf Confirmation}(\sigma,\text{msg})$: Through the protocol, the signer S can prove the validity of the signature $\sigma$ to the verifier V. This interactive process returns 1 if the signature is confirmed as valid; otherwise, it returns 0.
   \item $(0/1) \leftarrow {\sf Disavowal}(\sigma, \text{msg})$: This protocol is designed to allow a verifier to determine whether a given signature is invalid. The interactive process will return a value of 1 if the signature is determined to be invalid, and 0 if it is valid.
\end{itemize}

\end{enumerate}
\label{Def:undeniable}
\end{definition}

\noindent \textbf{Security Properties for Undeniable Signature Scheme}: The confirmation and disavowal protocols must be {\sf complete}, {\sf sound}, and {\sf zero-knowledge}. The entire protocol should satisfy {\sf unforgeability} and {\sf invisibility}. We follow the security properties from  \cite{rawal2022lattice, jao2014isogeny}.
\begin{itemize}
    \item[] {\sf Completeness}: If the signature $\sigma$ is valid then the probability of a successful confirmation for any message $msg$ with a valid signature $\sigma$ is 1.
    Similarly, if $\sigma$ is found to be invalid, the probability of a successful disavowal is also 1.
    \item[] {\sf Soundness}: In the case where the signer S is acting dishonestly, the probability of a successful confirmation when processing any message $msg$ with an invalid signature $\sigma$ is negligible. Similarly, if $\sigma$ is determined to be valid for a dishonest Signer S, the probability of a successful disavowal is also negligible.
    \item[] {\sf Zero-knowledge}: If the signature \(\sigma\) is valid, the verifier gains no knowledge about the secret other than the fact that the statement is true.
    \item[] {\sf Unforgeability}: The notion of unforgeability in the context of an undeniable signature scheme is defined through a game played by a challenger ($\mathcal{C}$) and an adversary ($\mathcal{A}$).
    %concept of unforgeability in the context of an undeniable signature scheme is defined through a game played between a challenger ($\mathcal{C}$) and an adversary ($\mathcal{A}$). 
    \begin{enumerate}
        \item The challenger $\mathcal{C}$ runs the $\text{KeyGen}$ algorithm and provides the adversary $\mathcal{A}$ with the verification key $\text{pk}$.
        \item $\mathcal{A}$ is allowed to make adaptive signing oracle queries for $i=1,\ldots, q$ for messages ${msg}_i$. For each signing query, $\mathcal{C}$ responds with a signature $\sigma_i$ computed using the $\text{Sign}$ algorithm. Additionally, $\mathcal{A}$ can access the confirmation/disavowal protocol for each received signature $\sigma_i$.
        \item Finally, $\mathcal{A}$ responds with a forge message-signature pair $({msg}^{*}, \sigma^{*})$.
        \item $\mathcal{A}$ wins if $({msg}^{*}, \sigma^{*})$ is valid and not equal to the messages queried before in the singing query phase.
    \end{enumerate}
    The scheme is considered {\em existentially unforgeable} under the chosen message attack (EUF-CMA) if the probability that any polynomial-time adversary $\mathcal{A}$ succeeds in the aforementioned scenario is negligible.
    %The signature scheme achieves {\em existential unforgeablility} under the chosen message attack (EUF-CMA) if for any polynomial-time adversary $\mathcal{A}$, the probability of success of $\mathcal{A}$ in the above game is negligible.
    
    \item[] {\sf Invisibility}: This property can be described by a game between a challenger $\mathcal{C}$ and an adversary $\mathcal{A}$.
    \begin{enumerate}
        \item The challenger $\mathcal{C}$ runs the $\text{KeyGen}$ algorithm and provides the adversary $\mathcal{A}$ with the verification key $\text{pk}$.
        \item $\mathcal{A}$ can issue a series of signing oracle queries for some message ${msg}_i$ and receives some signature $\sigma_i$. $\mathcal{A}$ is allowed to access the confirmation/disavowal oracle for each received signature $\sigma_i$.
        \item Eventually, $\mathcal{A}$ selects a message ${msg}^*$ and forwards it to $\mathcal{C}$.
        \item $\mathcal{C}$ selects $b \in \{0,1\}$ randomly. For $b=1$, it computes the {\em real} signature $\sigma^*$ for ${msg}^*$ using the secret key $\text{sk}$. Otherwise, it outputs a fake signature and sets it as $\sigma^*$ for ${msg}^*$.
        \item $\mathcal{A}$ can continue singing oracle queries. Here $\mathcal{A}$ do not have access to the confirmation/Disavowal oracle.
        \item Finally, $\mathcal{A}$ outputs a bit $b' \in \{0,1\}$.
        \item If $b=b'$, then $\mathcal{A}$ wins.
    \end{enumerate}
    The signature scheme attains the {\em invisibility} property if the probability that any polynomial-time adversary $\mathcal{A}$ succeeds in the previously described game is negligible.
    
    %if for any polynomial-time adversary $\mathcal{A}$, the probability of success of $\mathcal{A}$ in the above game is negligible.
    
\end{itemize}

%%%%%%%%%%%%%%%%%%%%%%%%%%%%%%%%%%%%%%%%%%%%%%%%%%%%%%%%%%%%%%%%%%%%%%%%%%%%%%%%%%%%%%%%%%%%%%%%%%%%%%%%%%%%%%%%%%%%%%%%%%%%%%%%%%%
%In this section we discuss the Module lattices, hardness assumptions on Module lattices, 
\subsection{Module Lattice}

 A module lattice \cite{bert2021implementation} with a polynomial structure is a type of lattice in which the basis vectors are polynomials with integer coefficients. This lattice naturally arises in the context of polynomial ring extensions, where it is constructed over a polynomial ring. The polynomial rings are defined as $\mathcal{R} = \mathbb{Z}[X]/(X^n + 1)$ and $\mathcal{R}_q = \mathbb{Z}_q[X]/(X^n + 1)$, where $q$ is a prime number and $n$ is an integer power of $2$.
 
\noindent\textbf{Ideal Lattice:} Let $ \mathcal{R} = \mathbb{Z}[X]/(f)$ be a quotient ring, where $f$ is a degree $n$ irreducible polynomial over $\mathbb{Z}[X]$. Ideal lattices correspond to the ideals of the ring $\mathcal{R}$. Notably, the quotient ring $\mathcal{R}$ is isomorphic to the ring $\mathbb{Z}^n$ with each polynomial mapped to an $n$-dimensional vector composed of its coefficients. Given an ideal $\mathcal{I}$ of $ \mathcal{R}$, the sublattice of $\mathbb{Z}^n$ that corresponds to $\mathcal{I}$ is called an ideal lattice. In short,  if $B$ is a basis of an ideal lattice, then $\mathcal{L}(B) \subset \mathbb{Z}^n$.
%by mapping a polynomial to an $n$-dimensional vector formed by its coefficients.When we take an ideal $\mathcal{I}$ of $ \mathcal{R}$, then the sublattice of $\mathbb{Z}^n$ corresponding to $\mathcal{I}$ is referred to as an ideal lattice. In summary,

\noindent\textbf{Module Lattice:} Module lattices are sublattices of $\mathcal{R}^k$ for some positive integer $k$. It is important to note that if $\mathcal{R} = \mathbb{Z}[X]/(X^n + 1)$, then $\mathcal{R}^k$ is isomorphic to $\mathbb{Z}^{kn}$. The module versions of {\sf Ring-SIS} and {\sf Ring-LWE} were first presented in \cite{langlois2015worst}.
%The module variants of {\sf Ring-SIS} and {\sf Ring-LWE} were introduced in \cite{langlois2015worst}.

\begin{definition}{Shortest Integer Solution Problem for Module Lattice} \\
{\sf Module-SIS}$_{l,k,q,\gamma}$: Given a matrix $\mathbf{A} \in \mathcal{R}_q^{l \times k}$, an integer q, and a real number $\gamma$, the goal is to find a non-zero vector $\mathbf{s} \in \mathcal{R}^k$ such that $\mathbf{A} \mathbf{s} = \mathbf{0} \mod q$ and $||\mathbf{s}|| \leq \gamma$.
\end{definition}

\begin{definition}{Decisional Learning with Errors for Module Lattices}\\ 
{\sf Decision~Module-LWE}$_{l,k,q,\gamma}$:
Consider a matrix $\mathbf{A} \in \mathcal{R}_q^{k \times l}$, an integer q, and a real number $\gamma$. Let $\mathbf{b} = \mathbf{A} \mathbf{s} + \mathbf{e} \mod{q}$, where $\mathbf{s}$ is chosen uniformly at random from $\mathcal{R}_q^l$ and  $\mathbf{e}$ is sampled from the discrete Gaussian distribution $D_{\mathcal{R}^k, \gamma}$. The problem is to distinguish between the distribution $(\mathbf{A}, \mathbf{b})$ and the uniform distribution over $\mathcal{R}_q^{k \times l} \times \mathcal{R}_q^k$.
\end{definition}

\subsection{Building Trapdoor Function for Module Lattice}
\label{trapmod}

In this section, we outline the procedure for constructing a trapdoor function on module lattices, based on the works of \cite{bert2021implementation, micciancio2012trapdoors}. We recall the representations of $k$-dimensional $q$-array lattice $\mathcal{L}^{\perp}(\mathbf{A})$ and its coset $\mathcal{L}^{\mathbf{u}}(\mathbf{A})$ for a matrix $\mathbf{A} \in \mathcal{R}^{n\times k}$ at first.
\begin{align*}
      \mathcal{L}^{\perp}(\mathbf{A})=\{\mathbf{x}\in \mathcal{R}^k : \mathbf{A}\mathbf{x}= \mathbf{0} \mod q\}.
\end{align*}
\begin{align*}
      \mathcal{L}^{\mathbf{u}}(\mathbf{A})=\{\mathbf{x}\in \mathcal{R}^k : \mathbf{A}\mathbf{x}=\mathbf{u} \mod q\}.
\end{align*}

\noindent \textbf{Ring G-Lattice:} The ring gadget vector is defined by,
\begin{align*}
\mathbf{y}^T = [1~~a~~a^2~\ldots~a^{m-1}] \in \mathcal{R}_q^{1\times m}, ~~~ {\text{where}}~~ m=\lceil \log_a q \rceil.
\end{align*}
The {\em Ring G-Lattice} is the $q$-array lattice,
\begin{align*}
\mathcal{L}^{\perp}(\mathbf{y}^T)=\{\mathbf{x}\in \mathcal{R}^m : \mathbf{y}^T\mathbf{x}=\mathbf{0} \mod q\}.
\end{align*}

\noindent \textbf{Module G-Lattice:} Using the ring gadget vector $\mathbf{y}^T$ we can construct a $G$-matrix as follows,
\begin{equation*}
\mathbf{G} = \mathbf{I}_l \otimes \mathbf{y}^T=
\begin{pmatrix}
\mathbf{y}^T &  & &\\
 & \mathbf{y}^T &  &\\
&  & \ddots & \\
 &  & &\mathbf{y}^T 
\end{pmatrix}
\in \mathcal{R}_q^{l \times {lm}}
\end{equation*}
where $\otimes$ is the tensor product between two vectors.
The {\em module G-lattice} is defined by,
\begin{align*}
\mathcal{L}^{\perp}(\mathbf{G})=\{\mathbf{x}\in \mathcal{R}^{lm} : \mathbf{G}\mathbf{x}=\mathbf{0} \mod q\}.
\end{align*}
\noindent \textbf{Trapdoor:} A trapdoor $\mathbf{T} \in \mathcal{R}^{(k-lm)\times lm}$ for the given matrix $\mathbf{A} \in \mathcal{R}_q^{l\times k}$ can be defined as,
\begin{equation*}
\mathbf{A} \left[\frac{\mathbf{T}}{\mathbf{I}_{lm}}\right]= \mathbf{L}\mathbf{G}
\end{equation*}
for some invertible matrix $\mathbf{L} \in \mathcal{R}_q^{l\times l}$, where $k=l(m+2)$. $\mathbf{I}_{lm}$ is the identity matrix of order $lm$.
%%%%%%%%%%%%%%%%%%%%%%%%%%%%%%%%%%%%%%%%%%%%%%%%
\begin{table}[ht]
    \centering
    \begin{tabular}{c}
  \toprule
       \textbf{Algorithm 1} {\sf MLTrapGen}$(\mathbf{L}\in \mathcal{R}_q^{l\times l},\beta > 0)$ to generate Trapdoor \\ \midrule
       1: \textbf{function} {\sf MLTrapGen}$(\mathbf{L}\in \mathcal{R}_q^{l\times l},\beta > 0)$  \hfill ~\\
       2: $\mathbf{B} \gets \mathcal{U}(\mathcal{R}_q^{l \times l})$ \hfill $\rhd \mathbf{B} \in \mathcal{R}_q^{l \times l}$\\
       3: $\mathbf{B}^{\prime} \gets \left[\mathbf{I}_l \mid \mathbf{B}\right]$ \hfill $\rhd \mathbf{B}^{\prime} \in \mathcal{R}_q^{l \times 2l}$\\
       4: $\mathbf{T} \gets D_{\mathcal{R}^{2l \times lm}, \beta}$ \hfill $\rhd \mathbf{T} \in \mathcal{R}^{2l \times lm}$\\
       5: $\mathbf{A} \gets \left[ \mathbf{B}^{\prime} \mid \mathbf{L}\mathbf{G} - \mathbf{B}^{\prime} \mathbf{T}\right]$ \hfill $\rhd \mathbf{A} \in \mathcal{R}_q^{l \times k}$\\
       6: \textbf{return} $(\mathbf{A}, \mathbf{T})$ \hfill~\\ \bottomrule
    \end{tabular}
    \label{tab:my_label_1}
\end{table}
%%%%%%%%%%%%%%%%%%%%%%%%%%%%%%%%%%%%%%%%

\begin{table*}[ht]

    \centering
    \scalebox{1.3}{
    \begin{tabular}{c}
    \toprule
    \textbf{Algorithm 2} {\sf MLSamplePre}($\mathbf{A}$, $\mathbf{T}$, $\mathbf{L}$, $\mathbf{w}\in \mathcal{R}^l$, $\chi$, $\alpha$) for sampling Gaussian preimage \\ 
    \midrule
    1: \textbf{function} {\sf MLSamplePre}($\mathbf{A}\in \mathcal{R}^{l \times k}$, $\mathbf{T} \in \mathcal{R}^{2l \times{lm}}$, $\mathbf{L} \in \mathcal{R}_{q}^{l \times l}$, $\mathbf{w}\in \mathcal{R}^l$, $\chi$, $\alpha$)   
 \hfill ~\\
2: $ \mathbf{p} \gets \sf {SamplePerturb} (\mathbf{T}, \chi, \alpha)$  \hfill  $\rhd \mathbf{p} \in \mathcal{R}^k$ \\
3: $\mathbf{v} \gets \mathbf{L}^{-1}(\mathbf{w}- \mathbf{A}\mathbf{p})$  \hfill  $\rhd \mathbf{v} \in \mathcal{R}^l$\\
4: $\mathbf{d} \gets D_{\mathcal{L}^\mathbf{v} (\mathbf{G}), \alpha}$  \hfill $\rhd \mathbf{d} \in \mathcal{R}^{lm}$\\
5: $\mathbf{u} \gets  \mathbf{p} + \left[\frac{\mathbf{T}}{\mathbf{I}}\right] \mathbf{d}$  \hfill $\rhd \mathbf{u} \in \mathcal{R}^k$\\
6: \textbf{return } $\mathbf{u}$ \hfill ~ \\ \bottomrule
    \end{tabular}}
    
    \label{tab:my_label}
\end{table*}

\begin{lemma}{\cite{bert2021implementation}}
Let $k = l(m + 2)$ with $m = \lceil \log_a q \rceil$ and q and l be positive integers. From the above algorithm, we have ${\sf{MLTrapGen}}(\mathbf{L} \in \mathcal{R}_q^{l \times l}, \beta > 0) \rightarrow (\mathbf{A}, \mathbf{T})$. Then $\mathbf{A}$ is computationally indistinguishable from uniform distribution (over $\mathcal{R}_q^{l\times k}$) based on the hardness assumption of {\sf Decision~Module-LWE}$_{l,q,\beta}$.
\end{lemma}

\subsection{Gaussian Preimage Sampling}
We discuss the Gaussian preimage sampling technique for module lattices, referred to as {\sf MLSamplePre} \cite{bert2021implementation}. This sampling technique consists of two underlying algorithms: \textbf{G-sampling} and \textbf{Perturbation sampling}.
\vspace{0.2cm}\\
\noindent\textbf{G-sampling:} The primary objective of G-sampling is to take a vector $\mathbf{w} \in \mathcal{R}^l$ and output a vector $\mathbf{u} \in \mathcal{R}^k$ from $\mathcal{L}^{\mathbf{w}}(\mathbf{G})$ by following a discrete Gaussian distribution with Gaussian parameter $\alpha$, where $\mathbf{G}$ is the G-matrix defined in Subsection \ref{trapmod}. The construction is detailed in \cite{genise2018faster, bert2021implementation}. Specifically, let $\mathbf{v} = \mathbf{L}^{-1}\mathbf{w}$. We then sample {\bf{d}} from the discrete Gaussian distribution $D_{\mathcal{L}^{\mathbf{v}}(\mathbf{G}), \alpha}$ and output $\mathbf{u} = \left[\frac{\mathbf{T}}{\mathbf{I}}\right] \mathbf{d}$. However, this process may leak some information about the trapdoor $\mathbf{T}$, which is why the \textbf{Perturbation sampling} algorithm has been introduced.\\
\noindent\textbf{Perturbation sampling:} Perturbation sampling techniques are designed to produce random values that are statistically indistinguishable from truly random values. This means that an observer should be unable to differentiate between the generated values and genuine random values, even with substantial computational power or knowledge of the sampling algorithm. For module lattices \cite{bert2021implementation}, perturbation sampling involves drawing a vector from the Gaussian distribution over $\mathcal{R}^k$. We refer to this algorithm as {\sf{SamplePerturb}}, which takes the trapdoor {\bf{T}} and two parameters, $\chi$ and $\alpha$, as inputs. It outputs a perturbation vector $\mathbf{p} \in \mathcal{R}^k$ such that when {\bf{p}} is added to $\left[\frac{\mathbf{T}}{\mathbf{I}}\right] \mathbf{d}$, no information about $\mathbf{T}$ can be determined.\\
%%%%%%%%%%%%%%%%%%%%%%%%%%%%%%%%%%%%%
\section{Proposed Undeniable Signature Scheme on Module Lattice}\label{Proposed}
In this section, we present our proposed undeniable signature scheme based on module lattices. Our approach involves three main algorithms: {\sf MLKeyGen}, {\sf MLSign}, and {\sf MLVerification}. The {\sf MLVerification} algorithm includes two subprotocols: {\sf confirmation} and {\sf disavowal}. The public key-secret key pair $({\sf PK}, {\sf SK})$ can be generated by the signer using the {\sf MLKeyGen} algorithm. In the {\sf MLSign} algorithm, the signer produces a signature $\sigma$ for a given message $msg$. During the {\sf MLVerification} algorithm, the verifier checks the validity of the message-signature pair $(msg, \sigma)$ using either the {\sf confirmation} or the {\sf disavowal} protocol. A detailed explanation of each component is provided below:

\noindent \textbf{Parameters:} Let $n$ be a security parameter. $q= poly(n)$ and the size of $q$ is $m=\lceil\log_a q \rceil$. $\beta,\chi, \alpha$ are Gaussian parameters. $t,s,l,k$ are positive integers with $k=l(m+2)$. We need two cryptographically secure collision-resistance hash functions: $h:\{0,1\}^{*} \longrightarrow \mathcal{R}_q^l$ and $h_1:\{0,1\}^{*} \longrightarrow \mathcal{R}_{q}^{l \times k}$.

\begin{itemize}
\item {$({\sf PK}, {\sf SK}) \leftarrow$ {\sf MLKeyGen}($1^n$):} This randomized key generation algorithm runs by the signer.

\begin{enumerate}
    \item The signer runs the ${\sf  MLTrapGen}(\mathbf{I}, \beta)$ algorithm to generate a pair $(\mathbf{A}, \mathbf{T})$, where $\mathbf{A} \in \mathcal{R}_q^{l \times k}$,  $\mathbf{T} \in \mathcal{R}^{2l \times lm}$ and $\mathbf{I}$ is the identity matrix of order $l$.
    \item Chooses a seed ${\sf  SD}$ randomly from  $\{0,1\}^{*}$.
    \item Computes $h_1 ({\sf SD})=\mathbf{L} \in \mathcal{R}_q^{l \times k}$.
    \item Selects $\mathbf{v}$ from $D_{\mathcal{R}^k, \beta,}$ and computes $\mathbf{H} = \mathbf{L} \mathbf{v} \mod{q}$.
\end{enumerate}

This algorithm outputs a public key ${\sf PK}= (\mathbf{A}, {\sf SD}, \mathbf{H})$ and a secret key ${\sf SK} = (\mathbf{T}, \mathbf{v})$.

\item {$\sigma \leftarrow$ {\sf MLSign}($msg,{\sf SK}$):} This randomized algorithm generates a signature $\sigma$ for a given message. This algorithm runs by the signer.
\begin{enumerate}
\item Computes $h_1(msg)=\mathbf{M} \in \mathcal{R}_q^{l \times k}$.
\item Chooses $\mathbf{r}$ randomly from $\{0,1\}^{*}$.
\item Calculates $h(\mathbf{L} \parallel \mathbf{r})= \sigma_1 \in \mathcal{R}_q^{l}$.
\item Runs the ${\sf  MLSamplePre}$ algorithm and calculates ${\sf MLSamplePre} (\mathbf{A}, \mathbf{T},  \mathbf{I}, \sigma_1, \chi, \alpha)= \sigma_2$.
\item Generates $\sigma_3= \mathbf{M}\mathbf{v} \mod{q}$.

\end{enumerate}
The algorithm outputs a signature $\sigma=(\sigma_1, \sigma_2, \sigma_3)$ for the message $msg$.

\item {$(0/1) \leftarrow$ {\sf  MLVerification} ($msg,\sigma, {\sf PK}$):} This verification algorithm involves of two algorithms: the $\sf {Confirmation}$ and the $\sf {Disavowal}$ protocol. Before initiating those protocols, the verifier will first check $\sigma_1$ and $\sigma_2$ as follows:

\begin{enumerate}
\item[1.] Check whether $\|\sigma_2\| \leq t \chi \sqrt{kn}$.
\item[2.] Verifies $\mathbf{A} \sigma_2 = \sigma_1 \mod{q}$.
\end{enumerate}
In ${\sf  Confirmation}$ or ${\sf Disavowal}$ the verifier only proves the correctness of $\sigma_3$. Note that the verifier can compute $\mathbf{L}$ and $\mathbf{M}$ using the $\sf {PK}$ and the message $msg$ respectively.

\begin{itemize}
    
\item[] {${\sf Confirmation}/ {\sf Disavowal}$:} This protocol runs between the signer and the verifier as follows:
\begin{enumerate}
\item The signer randomly chooses $\mathbf{e}$ from $\mathcal{R}^k$ and a permutation $\phi$ over the set $\{1,\ldots,k\}$.
\item It computes $\mathbf{j}=\mathbf{M}\mathbf{e} \mod{q}$.
\item Evaluates three commitments $comm_1= h(\phi || (\mathbf{L}+\mathbf{M}) \mathbf{e} \mod{q})$, $comm_2= h(\phi(\mathbf{e}))$ and $comm_3= h(\phi(\mathbf{v}+\mathbf{e}))$.
\item The signer outputs $comm_1$, $comm_2$ and $comm_3 $ as commitments.
\item As a challenge the verifier outputs $ch \in \{0,1,2\}$.

\item If $ch=0$, the signer responses $\phi(\mathbf{v})$ and $\phi(\mathbf{e})$.\\
If $ch=1$, it responses $\phi$, $\mathbf{v}+\mathbf{e}$ and $\mathbf{j}$.\\
If $ch=2$, it responses $\phi$, $\mathbf{e}$.
 \item The verifier performs the verification as follows:
     \begin{enumerate}
     \item[i.] For $ch=0$ the verifier checks the correctness of $comm_2$ and $comm_3$.
     \item[ii.] 
     For $ch=1$, the verifier first checks if the condition $\mathbf{M}(\mathbf{v}+\mathbf{e})- \mathbf{j} = \sigma_3$ is satisfied. If this condition holds, the verifier proceeds with the following steps:
     \begin{itemize}
     \item[--] It checks the correctness of $comm_3$.
     \item[--] ${ \sf Confirmation}$: The interaction protocol is termed a ${\sf Confirmation}$ protocol if $comm_1 = h(\phi || (\mathbf{L}+\mathbf{M}) (\mathbf{v}+\mathbf{e})- \mathbf{H}- \sigma_3 \mod{q})$.
     \item[--] ${\sf Disavowal}$: If $comm_1 \neq h(\phi || (\mathbf{L}+\mathbf{M}) (\mathbf{v}+\mathbf{e})- \mathbf{H}- \sigma_3 \mod{q})$, it is referred to as a ${\sf Disavowal}$ protocol.
     \end{itemize}
     \item[iii.] Finally, for $ch=2$ the verifier checks the validity of $comm_1$ and $comm_2$.
     
     \end{enumerate}

\end{enumerate}

 The verifier outputs 1 after a successful verification otherwise outputs 0.
 \end{itemize}
\end{itemize}

\section{Security} \label{Security}
We analyse our proposed scheme's security in this part.  We demonstrate that the {\sf Confirmation} and {\sf Disavowal} protocols within our construction are both complete and sound, and uphold zero-knowledge. Additionally, we show that the protocol meets the {\sf unforgeability} and {\sf invisibility} criteria.
%%%%%%%%%%%%%%%%%%%%%%%%%%%%%%%%%%%
\subsection{Zero-Knowledge Proof of  {\sf Confirmation} and {\sf Disavowal} Protocol:}
\begin{theorem}
The {\sf Confirmation} and {\sf Disavowal} protocol in our proposed scheme are complete.
\end{theorem}
\begin{IEEEproof}
Let the signature $\sigma = (\sigma_1, \sigma_2, \sigma_3)$ be valid for a message $msg$. Hence, for any $\mathbf{e} \in \mathcal{R}^k$ and any permutation over the set $\{1, \ldots, k\}$, the following hold:
\begin{itemize}
    \item[--] $\mathbf{M}(\mathbf{v} + \mathbf{e}) - \mathbf{j} = \sigma_3$.
    \item[--] $comm_1 = h(\phi || (\mathbf{L} + \mathbf{M})(\mathbf{v} + \mathbf{e}) - \mathbf{H} - \sigma_3 \mod{q}) = h(\phi || (\mathbf{L} + \mathbf{M}) \mathbf{e} \mod{q})$.
    \item[--] $comm_3 = h(\phi(\mathbf{v} + \mathbf{e})) = h(\phi(\mathbf{v})) + h(\phi(\mathbf{e}))$.
\end{itemize}
Therefore, the $\sf {Confirmation}$ protocol executes successfully.

Now, let $\sigma_1$ and $\sigma_2$ be valid during verification, while $\sigma_3$ is invalid for a message $msg$. Then, for a chosen $\mathbf{e}$ and $\phi$, the following hold:
\begin{itemize}
    \item[--] $\mathbf{M}(\mathbf{v} + \mathbf{e}) - \mathbf{j} = \sigma_3$.
    \item[--] $comm_1 = h(\phi || (\mathbf{L} + \mathbf{M})(\mathbf{v} + \mathbf{e}) - \mathbf{H} - \sigma_3 \mod{q}) \neq h(\phi || (\mathbf{L} + \mathbf{M}) \mathbf{e} \mod{q})$.
    \item[--] $comm_3 = h(\phi(\mathbf{v} + \mathbf{e})) = h(\phi(\mathbf{v})) + h(\phi(\mathbf{e}))$.
\end{itemize}
Therefore, the $\sf {Disavowal}$ protocol executes successfully.
\end{IEEEproof}
%---------------------------------------

\begin{theorem}
The probability of successfully executing the {\sf Confirmation} protocol with an invalid $\sigma_3$ is negligible.
\end{theorem}

\begin{IEEEproof} We consider the scenario where $\sigma_3$ is invalid, i.e., $\sigma_3 \neq \mathbf{M}\mathbf{v} \mod{q}$. For the {\sf Confirmation} protocol to execute successfully, the verifier must validate the responses given by the signer to the corresponding challenges. The signer responds as follows:
\begin{enumerate}
\item If $ch=0$, the signer selects $\phi_1$ over the set $\{1,\ldots,k\}$, $\mathbf{e}_1\in \mathcal{R}^k$ and responses as $\phi_1(\mathbf{v})$ and $\phi_1(\mathbf{e}_1)$.
\item If $ch=1$, it selects $\phi_2$ over $\{1,\ldots,k\}$, $\mathbf{e}_2\in \mathcal{R}^k$ and responses as $\phi_2$, $\mathbf{v}+\mathbf{e}_2$ and $\mathbf{j}= \mathbf{M}\mathbf{e}_2 \mod{q}$.
\item If $ch=2$, it selects $\phi_3$ over $\{1,\ldots,k\}$, $\mathbf{e}_3\in \mathcal{R}^k$ and responses as $\phi_3$ and $\mathbf{e}_3$.
\end{enumerate}

We consider the following cases based on the challenge $ch$:
\begin{description}
\item[$ch=0$]: In this case, $comm_1 = h(\phi_1 || (\mathbf{L} + \mathbf{M}) \mathbf{e}_1 \mod{q})$, $comm_2 = h(\phi_1(\mathbf{e}_1))$, and $comm_3 = h(\phi_1(\mathbf{v} + \mathbf{e}_1))$. The verifier checks $comm_2$ and $comm_3$, which can be done perfectly.

\item[$ch=1$]: In this case, $comm_1 = h(\phi_2 || (\mathbf{L} + \mathbf{M}) \mathbf{e}_2 \mod{q})$, $comm_2 = h(\phi_2(\mathbf{e}_2))$, and $comm_3 = h(\phi_2(\mathbf{v} + \mathbf{e}_2))$. Here, the verifier checks $comm_1$ and $comm_3$. First, the verifier verifies if $\mathbf{M}(\mathbf{v} + \mathbf{e}_2) - \mathbf{j} = \sigma_3$ holds. Next, the verifier checks $comm_3$, which can also be verified without issue. For $comm_1$, the verifier checks if $h(\phi_2 || (\mathbf{L} + \mathbf{M})(\mathbf{v} + \mathbf{e}_2) - \mathbf{H} - \sigma_3 \mod{q})$ equals $comm_1$. However, $(\mathbf{L} + \mathbf{M})(\mathbf{v} + \mathbf{e}_2) - \mathbf{H} - \sigma_3 \mod{q} = (\mathbf{M} + \mathbf{L})\mathbf{e}_2 + \mathbf{M}\mathbf{v} - \sigma_3 \mod{q} \neq comm_1$ since $\sigma_3 \neq \mathbf{M}\mathbf{v} \mod{q}$. Thus, in this case, verification fails.

\item[$ch=2$]: In this case, $comm_1 = h(\phi_3 || (\mathbf{L} + \mathbf{M}) \mathbf{e}_3 \mod{q})$, $comm_2 = h(\phi_3(\mathbf{e}_3))$, and $comm_3 = h(\phi_3(\mathbf{v} + \mathbf{e}_3))$. Here, the verifier checks $comm_1$ and $comm_2$. With the given information, the verifier can compute both $comm_1$ and $comm_2$ correctly.
\end{description}

Hence, for one round, the probability of successfully completing the {\sf confirmation} protocol with an invalid $\sigma_3$ is $\frac{2}{3}$. If the protocol runs for $n$ rounds, the probability becomes $(\frac{2}{3})^n$, which is a negligible function of $n$.
\end{IEEEproof}

%---------------------------------------------

\begin{theorem}
The probability of successfully executing the $\sf {Disavowal}$ protocol is negligible if $\sigma_3$ is valid. 
\end{theorem}
\begin{IEEEproof}
We consider two scenarios: 1) for an honest signer and 2) for a dishonest signer to prove this theorem. Suppose a valid signature $\sigma_3$ is generated by an honest signer, i.e., $\sigma_3 = \mathbf{M}\mathbf{v} \mod{q}$. The verifier's task is to validate $\sigma_3$ through interactions with the signer. In the case of an honest signer, the responses are as follows:
\begin{enumerate}
\item If $ch=0$, the signer selects $\phi_1$ over $\{1,\ldots,k\}$, $\mathbf{e}_1\in \mathcal{R}^k$ and responses as $\phi_1(\mathbf{v})$ and $\phi_1(\mathbf{e}_1)$.
\item If $ch=1$, it selects $\phi_2$ over $\{1,\ldots,k\}$, $\mathbf{e}_2\in \mathcal{R}^k$ and responses as $\phi_2$, $\mathbf{v}+\mathbf{e}_2$ and  $\mathbf{j}= \mathbf{M}\mathbf{e}_2 \mod{q}$.
\item If $ch=2$, it selects $\phi_3$ over the set $\{1,\ldots,k\}$, $\mathbf{e}_3\in \mathcal{R}^k$ and responses as $\phi_3$ and $\mathbf{e}_3$.
\end{enumerate}

\noindent Verifier's Checks:
\begin{description}
\item[$ch=0$]: In this case, $comm_1= h(\phi_1 || (\mathbf{L}+\mathbf{M}) \mathbf{e}_1 \mod{q})$, $comm_2= h(\phi_1(\mathbf{e}_1))$ and $comm_3= h(\phi_1(\mathbf{v}+\mathbf{e}_1))$. The verifier has to check $comm_2$ and $comm_3$ which can be done perfectly.

\item[$ch=1$]: In this case, $comm_1= h(\phi_2 || (\mathbf{L}+\mathbf{M}) \mathbf{e}_2 \mod{q})$, $comm_2= h(\phi_2(\mathbf{e}_2))$ and $comm_3= h(\phi_2(\mathbf{v}+\mathbf{e}_2))$. The verifier checks if $\mathbf{M}(\mathbf{v} + \mathbf{e}_2) - \mathbf{j} = \sigma_3$. Since $\sigma_3$ is valid and the signer is honest, this holds easily. The verifier can check $comm_3$ without issue. To verify $comm_1$, the verifier checks whether $h(\phi_2 || (\mathbf{L} + \mathbf{M}) (\mathbf{v} + \mathbf{e}_2) - \mathbf{H} - \sigma_3 \mod{q})$ equals $comm_1$. Since $\sigma_3 = \mathbf{M}\mathbf{v} \mod{q}$, this verification is satisfied.

\item[$ch=2$]: In this case, $comm_1= h(\phi_3 || (\mathbf{L}+\mathbf{M}) \mathbf{e}_3 \mod{q})$, $comm_2= h(\phi_3(\mathbf{e}_3))$ and $comm_3= h(\phi_3(\mathbf{v}+\mathbf{e}_3))$. In this case the verifier has to check $comm_1$ and $comm_2$. With the given information the verifier can compute $comm_1$ and $comm_2$ correctly.
\end{description}
Consequently, for an honest signer, the probability of the {\sf Disavowal} protocol running if $\sigma_3$ is legitimate is zero.

\noindent Now, let us consider a scenario where a dishonest signer possesses a valid signature $\sigma$ and attempts to run the {\sf Disavowal} protocol. The signer tries to deceive the verifier into believing that the signature $\sigma$ is invalid. To do this, during the interaction phase, the signer selects a vector $\mathbf{v}^{\prime}$ randomly (distinct from the original vector $\mathbf{v}$) and chooses some random vectors to present challenges and their corresponding responses to the verifier.

In the case of $ch=1$, note that the relation $\mathbf{M}(\mathbf{v}^{\prime} + \mathbf{e}_2) - \mathbf{j} = \sigma_3$ does not hold because the signer is dishonest, and $\sigma_3$ is valid. Therefore, the dishonest signer cannot successfully run the {\sf Disavowal} protocol, even with a valid $\sigma_3$.

\end{IEEEproof}
%-----------------------------

\begin{theorem}
The {\sf Confirmation} and {\sf Disavowal} protocols attain statistical zero-knowledge properties.
\end{theorem}

\begin{IEEEproof}
We construct a simulator $\mathcal{S}$ that, using public information, generates a simulated transcript to prove the desired property. Initially, $\mathcal{S}$ randomly selects $\overline{ch} \in \{0,1,2\}$, representing a prediction of the value that the cheating verifier $\mathcal{CV}$ will not choose before interacting with $\mathcal{CV}$. We will now discuss each possible case.\\

\noindent $\overline{ch}=0$: $\mathcal{S}$ computes a vector $\mathbf{v}^{\prime} \in \mathcal{R}^k$ such that $\mathbf{H} + \sigma_3 = (\mathbf{L} + \mathbf{M})\mathbf{v}^{\prime} \mod q$. Next, it randomly selects a permutation $\phi^{\prime}$ from the set $\{1, \dots, k\}$ and a vector $\mathbf{e}^{\prime} \in \mathcal{R}^k$. Finally, $\mathcal{S}$ computes $\mathbf{j}^{\prime} = \mathbf{M} \mathbf{e}^{\prime} \mod q$. The corresponding commitments are as follows:
\begin{itemize}
    \item $comm^{\prime}_1= h(\phi^{\prime} || (\mathbf{L}+\mathbf{M}) \mathbf{e}^{\prime} \mod{q})$
    \item $comm^{\prime}_2= h(\phi^{\prime}(\mathbf{e}^{\prime}))$
    \item $comm^{\prime}_3= h(\phi^{\prime}(\mathbf{v}^{\prime}+\mathbf{e}^{\prime}))$
\end{itemize}
$\mathcal{S}$ sends all commitments to $\mathcal{CV}$. Then, $\mathcal{CV}$ outputs a challenge $ch \in \{0, 1, 2\}$. The subsequent responses from $\mathcal{S}$ are:
\begin{enumerate}
\item For $ch=0$, $\mathcal{S}$ outputs $\perp$ and aborts.
\item For $ch=1$, $\mathcal{S}$ outputs $(comm^{\prime}_1, comm^{\prime}_2, comm^{\prime}_3, \phi^{\prime}, \mathbf{v}^{\prime}+\mathbf{e}^{\prime}, \mathbf{j}^{\prime})$.
\item For $ch=2$, $\mathcal{S}$ outputs $(comm^{\prime}_1, comm^{\prime}_2, comm^{\prime}_3, \phi^{\prime}, \mathbf{e}^{\prime})$.
\end{enumerate}
When $ch=1$, the real transcript would be $(comm_1, comm_2, comm_3, \phi, \mathbf{v}+\mathbf{e}, \mathbf{j}^\prime)$. We assume that 
$(\phi^{\prime}, \mathbf{e}^{\prime})= (\phi, \mathbf{v}+\mathbf{e}-\mathbf{v}^{\prime})$. Therefore $(\mathbf{L}+\mathbf{M})(\mathbf{v}+\mathbf{e}-\mathbf{v}^{\prime})= (\mathbf{L}+\mathbf{M})\mathbf{e}$ and we can say that, $comm^{\prime}_1 = comm_1$ and $comm^{\prime}_3 = comm_3$. Hence, we can infer that the distributions of the  real transcript and the simulated transcript are statistically similar. A similar proof applies for the case when $ch = 2$.\\

\noindent $\overline{ch}=1$: $\mathcal{S}$ randomly selects a permutation $\phi^{\prime}$ from the set $\{1, \dots, k\}$, along with two vectors $\mathbf{e}^{\prime}$ and $\mathbf{v}^{\prime}$ from $\mathcal{R}^k$. The corresponding commitments are:
\begin{itemize}
    \item $comm^{\prime}_1= h(\phi^{\prime} || (\mathbf{L}+\mathbf{M}) \mathbf{e}^{\prime} \mod{q})$
    \item $comm^{\prime}_2= h(\phi^{\prime}(\mathbf{e}^{\prime}))$
    \item $comm^{\prime}_3= h(\phi^{\prime}(\mathbf{v}^{\prime}+\mathbf{e}^{\prime}))$
\end{itemize}
$\mathcal{S}$ sends them to $\mathcal{CV}$. Now $\mathcal{CV}$ outputs $ch \in \{0,1,2\}$ as a challenge. The subsequent responses by $\mathcal{S}$ are,
\begin{enumerate}
\item For $ch=0$, $\mathcal{S}$  outputs $(comm^{\prime}_1, comm^{\prime}_2, comm^{\prime}_3, \phi^{\prime}(\mathbf{v}^{\prime}), \phi^{\prime}(\mathbf{e}^{\prime}))$.
\item For $ch=1$, $\mathcal{S}$ returns $\perp$ and halts.
\item For $ch=2$, $\mathcal{S}$ outputs $(comm^{\prime}_1, comm^{\prime}_2, comm^{\prime}_3, \phi^{\prime}, \mathbf{e}^{\prime})$.
\end{enumerate}
When $ch=0$, the real transcript would be $(comm_1, comm_2, comm_3,  \phi(\mathbf{v}), \phi(\mathbf{e}))$. Let $\psi$ be a permutation over $\{1,\ldots, k\}$ with $\psi(x)=x^\prime$. We set $\phi^{\prime} = \phi \circ \psi^{-1}$ and $\mathbf{e}^{\prime} = \psi (\mathbf{e})$. Then we have, $\phi^{\prime}(\mathbf{e}^\prime) =  \phi (\mathbf{e})$ and $\phi^{\prime}(\mathbf{v}^\prime) =  \phi (\mathbf{v})$. Thus, we can say that the distributions of the real transcript and the simulated transcript are statistically comparable. The case for $ch = 2$ is straightforward.\\

\noindent $\overline{ch}=2$: We omit the discussion as it can be addressed in a similar manner.

Therefore, we have constructed a simulator $\mathcal{S}$ that can simulate the real transcript with probability $\frac{1}{3} + {\sf negl}(n)$.
\end{IEEEproof}

\subsection{Overall Security of Our Proposed Signature Scheme}

\begin{theorem}
Our proposed undeniable signature scheme achieves existential unforgeability under a chosen-message attack (EUF-CMA) security in the random oracle model under the hardness of ${\sf Module-SIS}_{l,k,q,\gamma}$ and ${\sf Decision~Module-LWE}_{l,k,q,\beta}$ where $\gamma = 2 t \chi \sqrt{kn}$.
\end{theorem}
\begin{IEEEproof}
We will prove this theorem by contradiction. Suppose there exists an adversary $\mathcal{A}$ that can break the existential unforgeability of the proposed undeniable signature scheme with non-negligible probability. If this is true, then it would imply the ability to solve both the ${\sf Decision~Module-LWE}_{l,k,q,\beta}$ and ${\sf Module-SIS}_{l,k,q,\gamma}$ problems, where $\gamma = 2t \chi \sqrt{kn}$. The adversary $\mathcal{A}$ is given the public key ({\sf PK}) and is allowed to make a polynomial number of queries to the hash oracle and the signature oracle before producing a forged message-signature pair.

\begin{enumerate}
\item If the pairs $({\sf SD}, \mathbf{L})$ and $(msg, \mathbf{M})$ already exist in the database, the challenger returns the same values for $h$ and $h_1$, respectively. Otherwise, it generates new outputs using the hash functions and stores them in the database.

\item Whenever $\mathcal{A}$ submits a sign query for a message ${msg}^{\prime}$, the challenger first checks if the corresponding signature $\sigma_{{msg}^\prime}$ is already in the database. If so, the challenger responds with $\sigma_{{msg}^\prime}$. Otherwise, the signature is generated by executing the {\sf MLSign} algorithm and returned along with the corresponding signature.
\end{enumerate}

At some point $\mathcal{A}$ outputs a forge signature $\sigma^{*}$ for some message ${msg}^{*}$. The challenger parses the signature as $\sigma^{*}=(\sigma^{*}_1, \sigma^{*}_2, \sigma^{*}_3)$. Note that in the construction $\sigma^{*}_1$ and $\sigma^{*}_2$ are independent of the message. Therefore two cases can be happened:
\begin{itemize}
    \item $\sigma^{*}_1$ and $\sigma^{*}_2$ are different from the queried database. This would be a forgery on the standalone GPV signature scheme based on module lattice and it admits {\em unforgeability} under the hardness of ${\sf Module-SIS}_{l,k,q,\gamma}$ and ${\sf Decision~Module-LWE}_{l,k,q,\beta}$ (see \cite{bert2021implementation}).
    
    \item If $\sigma^{*}_1$ and $\sigma^{*}_2$ match a previously queried signature, then only $\sigma^{*}_3$ has been forged. In this case, $\sigma^{*}_3 = \mathbf{M}^{*} \mathbf{v}^{*} \mod{q}$ for some $\mathbf{v}^{*}$, where $\mathbf{M}^{*} = h_1({msg}^*)$. The original signature for the message ${msg}^{*}$ is $\sigma_3 = \mathbf{M}^{*} \mathbf{v} \mod{q}$. Thus, the difference between the signatures is $\sigma_3 - \sigma^{*}_3 = \mathbf{M}^{*}(\mathbf{v} - \mathbf{v}^{*}) \mod{q}$, where $\|\mathbf{v} - \mathbf{v}^{*}\| \leq 2 t \chi \sqrt{kn}$. This represents an instance of the inhomogeneous ${\sf Module-SIS}_{l,k,q,\gamma}$.
\end{itemize}
Therefore our scheme attains {\em existential unforgeablility} under the chosen message attack under the hardness of ${\sf Module-SIS}_{l,k,q,\gamma}$ and ${\sf Decision~Module-LWE}_{l,k,q,\beta}$ where $\gamma = 2 t \chi \sqrt{kn}$.

\end{IEEEproof}

\begin{theorem}
    Our proposed undeniable signature scheme satisfies the {\em invisibility} property.
\end{theorem}
\begin{IEEEproof}
    The signature $\sigma$ in our proposed scheme consists of three components: $\sigma_1$, $\sigma_2$, and $\sigma_3$. The challenger $\mathcal{C}$ first runs the {\sf MLKeyGen} algorithm and sends the public key ({\sf PK}) to the adversary $\mathcal{A}$. The adversary may then submit signing oracle queries for certain messages ${msg}_i$ and receive the corresponding signatures $\sigma_i$. Furthermore, the {\sf confirmation} and {\sf disavowal} oracles are accessible to $\mathcal{A}$. Eventually, $\mathcal{A}$ comes with a challenge message ${msg}^*$, and the challenger proceeds as follows
    \begin{itemize}
        \item The values of $\sigma_1$ and $\sigma_2$ are computed, as in the original undeniable signature scheme, using {\sf PK} and the specific parameter $\mathbf{T}$ from the secret key ({\sf SK}).
        \item A vector $\mathbf{v}^{\prime \prime} \neq \mathbf{v}$ is sampled from $D_{\mathcal{R}^k , \beta}$, and $\sigma_3^{\prime \prime} = \mathbf{M}^{*}\mathbf{v}^{\prime \prime} \mod q$ is computed, where $\mathbf{M}^{*} = h_1({msg}^*)$.
        \item A bit $b \in \{0,1\}$ is randomly chosen. If $b = 1$, the actual signature is produced; otherwise, a fake signature is generated.
    \end{itemize}
In this stage, neither the {\sf confirmation} nor the {\sf disavowal} oracles are accessible to $\mathcal{A}$. Therefore, while the adversary can verify $\sigma_1$ and $\sigma_2$, it cannot verify $\sigma_3^{\prime \prime}$. As both $\sigma_3$ (in the real signature) and $\sigma_3^{\prime \prime}$ are uniformly distributed, $\mathcal{A}$ is unable to determine whether $\sigma_3^{\prime \prime}$ is valid. As a result, it cannot correctly guess the value of $b$, ensuring that the proposed signature scheme achieves the {\em invisibility} property.     
\end{IEEEproof}
%------------------------------------------------------------------------------

\section{Efficiency} \label{complexity}

This section analyzes the computation and communication costs associated with our proposed scheme. The computed sizes of the signatures, public keys, and secret keys are shown in Table \ref{Key size}. The public key, denoted as {\sf PK}, consists of three components: $\mathbf{A}$, $\sf{SD}$, and $\mathbf{H}$. The size of $\sf{SD}$ is excluded from Table \ref{Key size} because it can be generated using a random bit generator \cite{Bucci2005}.

\begin{table}[H]
 	
\begin{center}
\caption{Sizes (in bits) of the signature, public key, and secret key}
 \begin{tabular}{||c|c||}
 \hline
 Parameter & Size (in bits)\\
 \hline\hline
  $\mathbf{A}$  & $nlk\log q$ \\
  \hline
  $\mathbf{H}$ & $nl \log q$ \\
  \hline
  {\sf PK} & $nlk\log q+nl \log q = nl(k+1) \log q$ \\
  \hline\hline
  $\mathbf{T}$ & $2l^2 m n \log (2\beta \sqrt{n 2 l^2 m})$\\
  \hline
  $\mathbf{v}$ & $nl(m+2) \log q$ bi\\
  \hline
  {\sf SK} & $2l^2 m n \log (2\beta \sqrt{n 2 l^2 m})+ nl(m+2) \log q$\\
  \hline\hline
  $\sigma_1$ & $nl \log q$\\
  \hline
  $\sigma_2$ & $nk \log (2t \chi \sqrt{kn})$\\
  \hline
  $\sigma_3$ & $nl \log q$\\
  \hline
  $\sigma = (\sigma_1,\sigma_2,\sigma_3)$ & $2nl \log q + nk \log (2t \chi \sqrt{kn})$\\
  \hline
   \end{tabular}
\end{center}
\label{Key size}
\end{table}
We implemented the scheme using the C programming language to evaluate the practical time complexity of our proposed design. The detailed hardware and software specifications for the experimental setup are provided in Table~\ref{specs}.

\begin{table}[H]
\caption{Specific experimental scenarios for the implementation}
	\begin{center}
            \scalebox{1.3}{
		\begin{tabular}{||c|c||}
			\hline
			Language & C \\
			\hline
			Processor & AMD Ryzen 5 \\
			\hline
			RAM & 8 GB  \\
			\hline
			CPU Frequency & 2.1 GHz \\
			\hline
                Number of Cores & 4 cores\\
                \hline
		\end{tabular}}
	\end{center}
	\label{specs}
\end{table}

In Table \ref{param}, five parameter sets are presented. Set I corresponds to the ring setting, while the remaining sets pertain to the module setting. Parameter sets I and V are adopted from the work of Bert et al. \cite{bert2021implementation}. Parameters I-IV offer 96-bit quantum security, while V provides 128-bit quantum security. The communication costs associated with these parameter sets are detailed in Table \ref{communication}. The overall performance of our signature scheme, evaluated using the selected parameter sets, is summarized in Table \ref{performance}.

\begin{table}[H]
    \centering
    \caption{Parameters for implementation}
    \scalebox{0.85}{
    \begin{tabular}{|m{0.9cm}|c|c|c|c|c|}
    \hline
    Parameter set & I & II & III & IV & V\\
    \hline
    \multicolumn{1}{|c|}{$q$} & 1073707009 & 1073707009 & 1073707009 & 1073738753 & 1073738753 \\
    \multicolumn{1}{|c|}{$k$} & 30 & 30 & 30 & 30 & 30 \\
    \multicolumn{1}{|c|}{$n$} & 1024 & 512 & 256 & 256 & 256 \\
    \multicolumn{1}{|c|}{$d$} & 1 & 2 & 4 & 4 & 5 \\
    \multicolumn{1}{|c|}{$\beta$} & 7.00 & 7.00 & 7.00 & 7.00 & 5.55\\
    \multicolumn{1}{|c|}{$\chi$} & 83832.0 & 83832.0 & 83832.0 & 83832.0 & 83290.0\\
    \multicolumn{1}{|c|}{$\alpha$} & 48.34 & 48.34 & 48.34 & 48.34 & 54.35\\
    \multicolumn{1}{|c|}{$t$} & 12 & 12 & 12 & 12 & 14 \\
    \hline
    \end{tabular}}
    \label{param}
\end{table}

%--------------------------------------------------

\begin{table}[H]
    \centering
    \caption{Communication cost for different parameter sets}
    \scalebox{1.3}{
    \begin{tabular}{|c|c|c|c|c|c|}
    \hline
    {Parameter set} & I & II & III & IV & V\\
    \hline
    \hline
    {\bf{A}}(KB) & 124 & 25 & 496 & 496 & 775\\
    {\bf{H}}(KB) & 4 & 4 & 4 & 4 & 5 \\
    {\sf PK}(KB) & 128 & 29 & 500 & 500 & 780\\
    \hline
    \hline
    {\bf{T}}(KB) & 240 & 480 & 960 & 960 & 1500\\
    {\bf{v}}(KB) & 124 & 124 & 124 & 124 & 124\\
    {\sf SK}(KB) & 364 & 604 & 1084 & 1084 & 1624\\
   \hline
   \hline
    $\sigma$(KB) & 136 & 136 & 136 & 136 & 170\\
    \hline 
    \end{tabular}}
    \label{communication}
\end{table}

%----------------------------
\begin{table*}[h]
  \centering
  \caption{CPU cycles and Time taken (ms) for {\sf MLKeyGen}, {\sf MLSign} and {\sf MLverification}}
  \renewcommand{\arraystretch}{1.2}
  \begin{tabular}{|p{2cm}|c|c|c|c|c|c|c|}
    \hline
    \multirow{2}{1cm}{Parameter Set} & \multicolumn{2}{c|}{\sf MLKeyGen} & \multicolumn{2}{c|}{\sf MLSign} & \multicolumn{2}{c|}{\sf MLverification}\\
    % \hline
    % \mathbf{Inactive Modes} & \mathbf{Description}\\
    \cline{2-7}
    & CPU Cycle & Time Taken &  CPU Cycle & Time Taken & CPU Cycle & Time Taken \\
    %\hhline{~--}
    \hline
    I & 93410898 & 44.29 & 52855509 & 25.12 & 16500015 & 7.86\\ \hline
    II & 103719840 & 49.39 & 72123429 & 34.35  & 1749862 & 8.31 \\ \hline
    III & 229843257 & 109.45 & 104722212 &  49.87  & 18750081 & 8.93\\ \hline
    IV & 200929911 & 95.21 & 107095863 & 51.00  & 17185392 & 8.18 \\ \hline
    V & 322669662 & 153.13 & 151808391 & 72.19 & 21853188 & 10.41 \\ \hline
  \end{tabular}
  \label{performance}
\end{table*}
%%%%%%%%%%%%%%%%%%%%%%%%%%%%%%%%%%%%%%%%%%%%%%%%%%55555555

\section{Conclusion} \label{Conclusion}
A post-quantum undeniable signature scheme is presented in this work, based on the hardness of the {\sf Module-SIS} and {\sf Module-LWE} problems. To the best of our knowledge, this is the first undeniable signature scheme based on module lattices. We provide comprehensive proofs for all associated security properties, ensuring the robustness of our approach. Furthermore, we have implemented our protocol in C, and tested it across various parameter sets. For a 128-bit security level, the time required by our scheme for key generation, signing, and verification is 153.13 ms, 72.19 ms, and 10.41 ms respectively. These results highlight the practicality of our scheme in real-world applications. Future work may explore more secure signature schemes with lower communication costs, which could enhance the efficiency and practical use of the post-quantum undeniable signature scheme.

% \section*{Acknowledgement} 

\bibliographystyle{plain}
\bibliography{ref1}

\end{document}